\documentclass[onecolumn,showpacs,preprintnumbers,amsmath,amssymb,aps]{revtex4}
\usepackage{epsfig}

 \normalsize

\newcommand{\Ga}{\Gamma}
\newcommand{\De}{\Delta}
\def\beq{\begin{equation}}
\def\lsim{\raise0.3ex\hbox{$\;<$\kern-0.75em\raise-1.1ex\hbox{$\sim\;$}}}
\def\gsim{\raise0.3ex\hbox{$\;>$\kern-0.75em\raise-1.1ex\hbox{$\sim\;$}}}
\def\eeq{\end{equation}}
\def\be{\begin{equation}}
\def\ee{\end{equation}}
\def\bea{\begin{eqnarray}}
\def\eea{\end{eqnarray}}

\def\nl{\nonumber \\}
\def\roughly#1{\mathrel{\raise.3ex\hbox
{$#1$\kern-.75em\lower1ex\hbox{$\sim$}}}}

\def\sss{\scriptscriptstyle}

\def\bra#1{\left\langle #1\right|}
\def\ket#1{\left| #1\right\rangle}
\newcommand{\bers}{\begin{eqnarray*}}
\newcommand{\eers}{\end{eqnarray*}}
\newcommand{\bt}{\begin{itemize}}
\newcommand{\et}{\end{itemize}}

\def\th1{\theta_{LR}}
\def\th2{\theta_{RL}}

\def\Frac#1#2{\frac{\displaystyle{#1}}{\displaystyle{#2}}}

\def\sss{\scriptscriptstyle}

\def\bd{B_d^0}
\def\bdbar{{\overline{B_d^0}}}
\def\bs{B_s^0}
\def\bsbar{{\overline{B_s^0}}}
\def\bq{B_q^0}
\def\bqbar{{\overline{B_q^0}}}
\def\ks{K_{\sss S}}

\def\prd#1#2#3{{ Phys.\ Rev.} {\bf D#1}, #3 (#2)}

\def\prl#1#2#3{{ Phys.\ Rev.\ Lett.} {\bf #1}, #3 (#2)}

\def \kstar{{\bar{K}^*}}

\def\BKstarmumu{\bdbar\to \kstar \mu^+ \mu^-}

\begin{document}
\begin{flushright}
UMISS-HEP-2010-04\\
[10mm]
\end{flushright}

\title{Like-sign dimuon charge asymmetry in Randall-Sundrum model}
\author{Alakabha Datta } \thanks{datta@phy.olemiss.edu}
\author{Murugeswaran Duraisamy } \thanks{duraism@phy.olemiss.edu}
\affiliation{Department of Physics and Astronomy, 108 Lewis Hall,
University of Mississippi, Oxford, MS 38677-1848, USA.}
\author{Shaaban Khalil} \thanks{skhalil@bue.edu.eg}
\affiliation{Center for Theoretical Physics at the
British University in Egypt, Sherouk City, Cairo 11837, Egypt.\\
Department of Mathematics, Ain Shams University, Faculty of
Science, Cairo, 11566, Egypt.}
\date{\today}

\begin{abstract}
We confirm that in order to account for the recent D\O\ result
of large like-sign dimuon charge asymmetry, a
considerable large new physics effect in $\Gamma_{12}^s$ is
required in addition to a large CP violating phase in $B_s
-\bar{B}_s$ mixing. In the Randall-Sundrum model of warped geometry,
where the fermion fields reside in the bulk, new sources of flavor
and CP violation are obtained. We analyze the like-sign dimuon
asymmetry in this class of model as an example of the desired new
physics. We show that the wrong-charge asymmetry, $a_{sl}^s$, which is related to the dimuon asymmetry, is significantly
altered compared to the standard model value. However, 
experimental limits from $\Delta M_s$, $\Delta\Gamma_s$ as well as $K$ mixing and electroweak corrections constrain it
 to be greater than a  $\sigma$ away from
its experimental average value. 
This model cannot fully account for the  D\O\  anomaly  due to its inability to generate
a sufficient new contribution to the width difference $\Gamma^s_{12}$, even though the model can generate large contribution to the mass difference  $M^s_{12}$.

\end{abstract}
\maketitle
\section{Introduction}
The B factories, BaBar and Belle, have firmly established the CKM
mechanism as the leading order contributor to CP violating
phenomena in the quark sector. New physics (NP) effects can add to
the leading order term producing deviations from the standard
model (SM) predictions. These deviations are expected to be more
pronounced in rare FCNC processes, as they are suppressed in the
SM. The Belle experiment is scheduled for an upgrade \cite{kino}
which will result in very precise results in $B$ decays. LHCb is
ready to take data and is expected to make many important
measurements in $b$ quark decays. These measurements may reveal the
presence of new physics.

In recent years, there have been several measurements of $B$
decays which differ from the predictions of the SM by $\sim
2\sigma$. For example, in $ B \to \pi K$, the SM has difficulty
accounting for all the experimental measurements \cite{
piKupdate}. The measured indirect (mixing-induced) CP asymmetry in
several $b \to s$ penguin decays is not found to be identical to
that in $\bd\to J/\psi\ks$ \cite{hfag}, counter to expectations of
the SM and could be providing hints for new
physics \cite{dattaRparity}.  The large transverse polarization in
some penguin dominated decays to light vector particles, like
$B\to\phi K^*$ \cite{phiK*}, are also somewhat difficult to
understand in the SM where naively one expects the transverse
polarization amplitudes to be suppressed. A further effect has recently been
seen in the lepton sector: in the exclusive decay $\BKstarmumu$,
the forward-backward asymmetry has been found to deviate somewhat
from the predictions of the SM \cite{Belle,BaBar}. Although this
disagreement is not statistically significant, the Belle
experiment itself claims this measurement shows a clear hint of
physics beyond the SM \cite{BellePR}. There are also other
measurements like the branching ratio of $B \to \tau \nu$ measured
at Belle which appear to be in conflict with SM expectation
\cite{belle_tau}.

Most discrepancies reported above have appeared in $b \to s $ transitions and so it is obvious that measurements in $B_s$ mixing will be crucial in testing the SM and finding evidence of new physics. In the SM, $\bs-\bsbar$ mixing is generated at loop level and is suppressed. Many new physics models can contribute to $\bs-\bsbar$ mixing and can cause measurable deviations from the SM. There are already measurements in the
$\bs-\bsbar$ system where the mass difference $ \Delta M_s$ and the width difference
$\Delta \Gamma_s$ between the two mass eigenstates have been measured.
Two other measurements in the $B_s$ system have generated enormous interest as
they do not appear to agree with the SM predictions.  The first measurement is the phase of $\bs- \bsbar$ mixing
which can be measured via indirect CP violation in
$\bar{B}_s \to J/\psi \phi$. The CDF
\cite{CDFpsiphi} and D\O\ \cite{D0psiphi} Collaborations have measured
indirect CP violation in $\bar{B}_s \to J/\psi \phi$. The experiments
measured
$S_{\psi\phi} = -2\beta_s$, and found \cite{hfag}
\beq
 \beta_s =0.41^{+0.18}_{-0.15} \quad   or \quad   1.16^{+0.15}_{-0.18}~ . 
\eeq
This disagrees with the SM prediction
\beq
\beta_s^{\rm SM} = 0.019 \pm 0.001.
\label{betasSM}
\eeq
Implications of this measurement for NP models have been analyzed \cite{dattapsiphi, lenz1}.

The second measurement was made recently in the $\bs-\bsbar$ system  when the D\O\ Collaboration  measured the like-sign dimuon charge
asymmetry with 6.1 $fb^{-1}$ of data \cite{Abaz:1005}. The following result was reported:%
\bea
\label{DO.Absl}
A^b_{sl} &=& -(9.57 \pm 2.51 (stat) \pm 1.46 (syst))\times 10^{-3}.
\label{dimuon_result}
\eea
The like-sign dimuon charge asymmetry $A^b_{sl}$ for semileptonic
decays of $b$ hadrons produced in $\bar{p}p$ collision is defined
as %
\be%
\label{eq1.Absl}
A^b_{sl} = \frac{N_b^{++} - N_b^{--}}{N_b^{++} + N_b^{--}},%
\ee %
where $N^{++}_b$ and $N^{--}_b$ are the number of events containing
two $b$ hadrons that decay semileptonically into $\mu^+ \mu^+ X$ and $\mu^- \mu^- X$, respectively. The semileptonic  decays of both $B_d$ and $B_s$ can contribute to $A^b_{sl}$. The relation between $A^b_{sl}$ and the "wrong-charge" asymmetries $a^d_{sl}$ and $a^b_{sl}$ is given by \cite{Grossman:2006ce}
\bea
\label{eq2.Absl}
A^b_{sl}&=& \Big(\frac{f_d z_d}{f_d z_d +f_s z_s }\Big) a^d_{sl}+\Big(\frac{f_s z_s}{f_d z_d +f_s z_s }\Big) a^s_{sl},
\eea
where $z_q =1/(1-y^2_q)-1/(1+x^2_q)$ (q = d, s).
Here $f_d$ and $f_s$ denote the  production fraction of $B_d$ and $B_s$, and
the quantities
$x_q $ and  $y_q $ are given as,
\bea
\label{xydef1}
x_{q} &\equiv&\frac{\Delta M_{q}}{ \Gamma_{q} },~~y_{q}\equiv\frac{\Delta \Gamma_{q}}{2 \Gamma_{q} },
\eea
where $\Delta M_{q}$ and $\Delta \Gamma_{q}$ are the mass and width differences in the $\bq- \bqbar$ system.
The semileptonic wrong-charge asymmetry $ a^q_{sl}$ is defined as
\bea
a^q_{sl} &=& \frac{\Gamma(\bar{B}_q\rightarrow \mu^+ X)-\Gamma(B_q\rightarrow \mu^- X)}{\Gamma(\bar{B}_q\rightarrow \mu^+ X)+\Gamma(B_q\rightarrow \mu^- X)}. \
\label{asl1}
\eea

%
Using the known values  $f_d= 0.323 \pm 0.037$, $f_s= 0.118 \pm
0.015$,  $x_d =0.774 \pm 0.008$, $y_d \approx 0$, $x_s=26.2\pm 0.5$ and
$y_s=0.0046\pm 0.027$ \cite{Abaz:1005}, \cite{pdg} , one can
rewrite Eq.~(\ref{eq2.Absl}) as \bea \label{eq3.Absl} A^b_{sl}&=&
(0.506 \pm 0.043) a_{sl}^d + (0.49 \pm 0.043) a_{sl}^s~. \eea The
SM predictions for the charge asymmetries are \cite{Abaz:1005}
\bea \label{eq2.aqslSM} a^d_{sl} &=& (-4.8^{+1.0}_{-1.2})\times
10^{-4},~~~ a^s_{sl}= (2.1\pm 0.6)\times 10^{-5}. \eea The SM
result for $A^b_{sl}$ can be obtained using Eqs.~(\ref{eq3.Absl}) and
(\ref{eq2.aqslSM}) as \bea \label{eq4.AbslSM} A^b_{sl}&=&
(-2.3^{+0.5}_{-0.6})\times 10^{-4}, \eea which is  about 3.2
$\sigma$ away from the value in Eq.~(\ref{dimuon_result}).

The SM prediction of the charge symmetry $a^{d}_{sl}$ in Eq.~(\ref{eq2.aqslSM}) is consistent with the observed value
$a^{d}_{sl} = −0.0047 \pm  0.0046 $ \cite{Abaz:1005, hfag},  within errors. In order  to obtain  D\O\ measurement for $A^{b}_{sl}$ in Eq.~(\ref{DO.Absl}) using the measured $a^{d}_{sl}$, the value of the charge symmetry $a^{s}_{sl}$ needs to be  \cite{Abaz:1005}
\bea
\label{assl.cal1}
a^{s}_{sl} &=& -(14.6 \pm 7.5) \times 10^{-3}.
\eea
This value is much larger than its SM prediction in Eq.~(\ref{eq2.aqslSM}).
 The  D\O\ direct measurement of $a^{s}_{sl}=-(1.7 \pm 9.1 (stat)^{+1.4}_{-1.5}(sys))10^{-3}$ \cite{Abazov:2009wg}, is consistent with the SM value in Eq.~(\ref{eq2.aqslSM}). An average value for $a^{s}_{sl}$ can be extracted by combining the   D\O\ and CDF \cite{CDF:9015} measurements as \cite{Dobrescu:2010rh}
\bea
\label{asls.avg}
(a^s_{sl})_{avg} &\approx& -(12.7 \pm 5.0)\times 10^{-3}.
\eea
This average  value of $a^s_{sl}$ is about 2.5$\sigma$ away from its SM value  in Eq.~\ref{eq2.aqslSM}.
A confirmation of this deviation would be an unambiguous evidence for new physics, and already interpretations of this result in terms of NP have been performed in various extensions of the SM \cite{Dobrescu:2010rh, ew2,D0_theory, lig}.

In this work we  consider the
warped extra dimension  Randall-Sundrum (RS) model \cite{RS}.
This model was proposed to solve the hierarchy problem in the SM and in this framework some of the flavor puzzles in the SM can be addressed in the split fermion scenario with the  fermions located at different points in the extra dimension \cite{Gherghetta:2000qt,RSflavor, ag}.

\par In this paper we  work out the contribution to the parameters $\Gamma^s_{12}$ and $M_{12}^s$ in the $\bs-\bsbar$ system
for the general case of NP with operators that are of the vector and/or axial vector types. The general formula that we derive can  be used for several extensions of the SM. Taking the  RS model as an example for new physics we  use our general expressions to  compute the contribution to  $ \Gamma_{12}^s$ and $M_{12}^s$.

The paper is organized as follows. In the first section, we present an overview of the phenomenology of the $\bs-\bsbar$ system including constraints on NP with present measurements. In the second section, we present the general expression for $\Gamma^s_{12}$ and $M^s_{12}$ for general new physics containing vector and /or axial vector operators. FCNC effects in the RS model with split fermions are discussed in the next section. The subsequent sections contain  our numerical results and conclusions.

\section{ Model independent analysis of $\bs-\bsbar$ Mixing}
In this section we will briefly review the phenomenology of the $B^0_q-\bar{B}^0_q$ system for q = s,d. The formalism for $B$ mixing is well known but we will review it here for completeness  and study the constraints on NP imposed by measurements in this system.

The $B^0_q$ and $\bar{B}^0_q$ states can mix in the presence of weak interactions. The resulting mass eigenstates can differ in their masses and  lifetimes.
In the $B_q-\bar{B}^0_q$ system, the time evolution of the general state is governed by the Schr\"{o}dinger equation

\bea
i \frac{d}{dt} \begin{pmatrix}B_q(t)  \\
\bar{B}_q(t)\end{pmatrix} &=& {\cal{H}}_{q} \begin{pmatrix}B_q(t)  \\
\bar{B}_q(t)\end{pmatrix},
\eea
where the  Hamiltonian ${\cal{H}}_{q} $ is given in terms of the $2 \times 2$ Hermitian  mass ($M_q$) and the decay width ($\Gamma_q$) matrices
\bea
{\cal{H}}_{q} & =& \Big(M_q-\frac{i}{2} \Gamma_q \Big)= \begin{bmatrix}
 M^q_{11} -\frac{i}{2}\Gamma^q_{11} &  M^q_{12}  -\frac{i}{2} \Gamma^q_{12}\\
 M^{q*}_{12}  -\frac{i}{2} \Gamma^{q*}_{12}&  M^q_{11} -\frac{i}{2}\Gamma^q_{11}
\end{bmatrix}.
\eea
The mass eigenstates  are the eigenvectors of ${\cal{H}}_{q} $. The eigenvectors with the  lightest  and  heaviest mass eigenvalues can be written as
\bea
\ket{B^L_q} &=&  p \ket{B_q} + q \ket{\bar{B}^0_q},~~
\ket{B^H_q} =  p \ket{B_q} -q \ket{\bar{B}^0_q}\,,
\label{hl}
\eea

with $|p|^2+|q|^2$ = 1. The masses and widths of these mass eigenstates are
\bea
M_q^{H,L} &=& M^q_{11} \pm Re[\sqrt{\Big( M^q_{12}  -\frac{i}{2} \Gamma^q_{12}\Big)\Big( M^{q*}_{12}  -\frac{i}{2} \Gamma^{q*}_{12}\Big)}],\nl
\Gamma_q^{H, L} &=& \Gamma^q_{11} \mp 2 Im[\sqrt{\Big( M^q_{12}  -\frac{i}{2} \Gamma^q_{12}\Big)\Big( M^{q*}_{12}  -\frac{i}{2} \Gamma^{q*}_{12}\Big)}].
\eea

One can now construct the following observables
\bea
M_q  & = & \frac{M^H_q   + M^L_q}{2}= M^q_{11},~~~~\Ga_q  = \frac{\Ga^H_q   + \Ga^L_q}{2} = \Ga^q_{11}, \nl
\Delta M_q &=& M^{H}_q-M^{L}_q=2 Re[\sqrt{\Big( M^q_{12}  -\frac{i}{2} \Gamma^q_{12}\Big)\Big( M^{q*}_{12}  -\frac{i}{2} \Gamma^{q*}_{12}\Big)}],\nl
\Delta \Gamma_q &=& \Gamma^{L}_q-\Gamma^{H}_q=4 Im[\sqrt{\Big( M^q_{12}  -\frac{i}{2} \Gamma^q_{12}\Big)\Big( M^{q*}_{12}  -\frac{i}{2} \Gamma^{q*}_{12}\Big)}].
\eea
The mass difference, the width difference and the parameters in the eigenvectors expression in Eq.~(\ref{hl}) can be written as
\bea
(\Delta M_q)^2-\frac{1}{4} (\Delta \Gamma_q)^2 &=& 4(| M^q_{12} |^2 -\frac{1}{4} |\Gamma^q_{12}|^2),\nl
\Delta M_q\Delta \Gamma_q &=& -4  Re[M^{q}_{12}  \Gamma^{q*}_{12}],\nl \Big(\frac{q}{p}\Big)_q &= &- \frac{2 M^{q*}_{12}-i \Gamma^{q*}_{12}}{\Delta M_{q}+\frac{i}{2} \Delta \Gamma_{q}}.
\eea
One usually defines the two dimensionless quantities
\bea
\label{xydef}
x_{B_q} &\equiv&\frac{\Delta M_{q}}{ \Gamma_{q} },~~y_{B_q}\equiv\frac{\Delta \Gamma_{q}}{2 \Gamma_{q} }.
\eea
Measurements indicate $y_{B_q} \sim O(10^{-2})$ while $x_{B_q} \sim 1$. These results model independently imply
\bea
\Delta \Gamma_{q} <<\Delta M_{q}.
\eea
Thus to a good approximation
\bea
\label{obserdef}
 \Delta M_q  & = & 2 | M^q_{12} |,~~~
\Delta \Gamma_q =- \frac{2Re[M^{q}_{12}  \Gamma^{q*}_{12}] }{| M^q_{12} |}
=2 |\Gamma_{12}|\cos{\phi_q},~~\Big(\frac{q}{p}\Big)_q= -\frac{ M^{q*}_{12}}{| M^q_{12} |}\Big(1-\frac{1}{2} Im[\frac{\Gamma^{q}_{12}}{M^{q}_{12}}]\Big)\,,
\label{m_gamma}
\eea
where
\bea
\frac{M_{12}}{\Gamma_{12}} &=& -\frac{|M_{12}|}{|\Gamma_{12}|}e^{i \phi_q},
~~
\phi_q = arg\Big(-\frac{M_{12}}{\Gamma_{12}}\Big).
\eea

The semileptonic wrong-charge asymmetry $ a^q_{sl}$ is defined as
\bea
a^q_{sl} &=& \frac{\Gamma(\bar{B}_q\rightarrow \mu^+ X)-\Gamma(B_q\rightarrow \mu^- X)}{\Gamma(\bar{B}_q\rightarrow \mu^+ X)+\Gamma(B_q\rightarrow \mu^- X)} = Im[\frac{\Gamma^q_{12}}{M^q_{12}}],\nl
a^q_{sl} &=& \frac{|\Gamma_{12}|}{|M_{12}|} \sin{\phi_q}=\frac{\Delta \Gamma_q}{\Delta M_q } \tan{\phi_q}.
\label{asl}
\eea

\begin{figure}[htb!]
\centering
 \includegraphics[width=7.5cm]{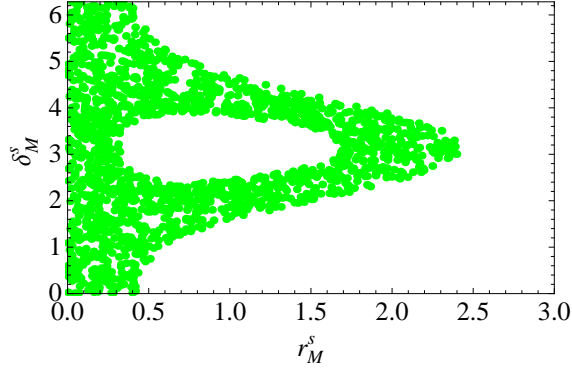}
\caption{ The allowed ranges for  $\delta^s_M-r^s_M$ from the combined result of CDF and  D\O\ in Eq.~(\ref{MBSexp}) are shown. }
\label{fig.rmdelmmass}
\end{figure}

The off-diagonal element $M^q_{12}$ of the  matrix M is related to the dispersive part of the $\Delta B=2$ transition amplitude
\bea
\label{M12def}
M^q_{12} &=& \frac{1}{2 m_{B_q}} \bra{\bar{B}^0_q} H^{\Delta B=2}_{eff}\ket{B^0_q},
\eea
 where $m_{B_q}$ is the $B_q$ meson mass. In the presence of NP contributions to $M^q_{12}$, it can be written as
\bea
\label{M12def2}
M^q_{12} &=& M^{q,SM}_{12} + M^{q,NP}_{12} = M^{q,SM}_{12} R^q_{M} e^{i \phi^q_{M}},
\eea
where
\bea
\label{M12def3}
R^q_{M}&= & |1+ r^q_{M} e^{i \delta^q_{M}}|= \sqrt{1+ 2 r^q_{M} \cos{\delta^q_{M}}+(r^{q}_{M})^2},~~\phi^q_{M}= arg[1+ r^q_{M} e^{i \delta^q_{M}}],
\eea
with $r^q_{M}e^{i \delta^q_{M}}=  M^{q,NP}_{12}/M^{q,SM}_{12}$. The mass difference in Eq.~(\ref{obserdef}) is modified
as
\bea
\label{massdifNP}
\Delta M_{q} = \Delta M^{SM}_{B_q} R^q_{M}.
\eea
The experimental result of $\Delta M_{{d}}$ is $\Delta M_{d} =
0.507 \pm 0.004~ {\rm ps}^{-1}$, which is consistent with the SM
expectation with $R^d_{M} \simeq 1$. This imposes a stringent constraint on any NP
contribution to the $B_d-\bar{B}_d$ mixing. One may estimate the
SM contribution to $\Delta M_{s}$ through the ratio $\Delta
M^{SM}_{s}/\Delta M^{SM}_{d}$, in order to minimize the
hadronic uncertainties. For quark mixing angle $\gamma
=67~^{\circ}$, one finds $\Delta M^{SM}_{s}\simeq 19~{\rm
ps}^{-1}$, which is in agreement with the latest measurements by
CDF \cite{Abulencia:2006ze} and  D\O\ \cite{Abazov:2006dm}:
\bea %
\label{MBSexp}
\Delta M_{s} &=& 17.77\pm 0.10 ({\rm stat.}) \pm 0.07({\rm syst.}), \nl
\Delta M_{s} &=& 18.53\pm 0.93 ({\rm stat.}) \pm 0.30({\rm
syst.}).
\eea %
 Fig. \ref{fig.rmdelmmass} shows the allowed ranges for  $\delta^s_{M}-r^s_{M}$ where we have neglected the SM phase in $M_{12}^s$. The green scatter points  satisfy the combined result of CDF and  D\O\ in Eq.~(\ref{MBSexp}) within the $1\sigma$ limit. The phase $\delta^s_{M}$ is varied in  the range [0, 2$\pi$], and it is not constrained below $r^s_{M} \lsim 0.4 $.  Also, one can conclude that
 $R^s_{M}$ is limited by: %
\bea
0.7 \lsim R^s_{M} \lsim 1.4 .%
\eea

The off-diagonal element $\Gamma^q_{12}$ of the  matrix $\Gamma$ is
related to the absorptive part of the transition  amplitude from
$B_q$ to $\bar{B}_q$. $\Gamma^q_{12}$ can be written as
\bea
\label{Gamdef1}
\Gamma^q_{12} &=&\frac{1}{2
m_{B_q}}\bra{\bar{B}_q}{\cal{T}}\ket{B_q}, \eea
where the transition operator ${\cal{T}}$ is defined by
\bea
\label{eq.trans1}
 {\cal{T}} &=& Im\Big[i \int d^4 x T
{\cal{H}}^{\Delta B=1}_{eff}(x) {\cal{H}}^{\Delta B=1}_{eff}(0)
\Big].
\eea
In the presence of NP contributions to $\Gamma^q_{12}$, it can be written as
\bea
\label{Gamdef2}
\Gamma^q_{12} &=& \Gamma^{q,SM}_{12} + \Gamma^{q,NP}_{12} = \Gamma^{q,SM}_{12} R^q_{\Gamma} e^{i \phi^q_{\Gamma}},
\eea
where
\bea
\label{Gamdef3}
R^q_{\Gamma}&= & |1+ r^q_{\Gamma} e^{i \delta^q_{\Gamma}}|= \sqrt{1+ 2 r^q_{\Gamma} \cos{\delta^q_{\Gamma}}+(r^q_{\Gamma_q})^2},~~\phi^q_{\Gamma}= arg[1+ r^q_{\Gamma} e^{i \delta^q_{\Gamma}}],
\eea
with $r^q_{\Gamma}e^{i \delta^q_{\Gamma}}= \Gamma^{NP}_{12}/\Gamma^{SM}_{12}$. The width difference in Eq.~\ref{obserdef} is modified
as
\bea
\label{GammadifNP}
\Delta \Gamma_{q} = 2 \vert \Gamma^{q,SM}_{12}\vert  R^q_{\Gamma} \cos{(\phi^q_{SM}+ \phi^q_{M}-\phi^q_{\Gamma})}.
\eea
The decay width difference $\De \Ga_s$ has been measured
independently.  The angular analysis of $\bar{B}_s \to J/\psi \phi$ gives
\cite{hfag,Dighe:1995pd,Dighe:1998vk}
\beq
 \Delta \Gamma_s = \pm (0.154^{+0.054}_{-0.070}) \; {\rm ps}^{-1} ~,
\label{width_diff_expt}
\eeq
to be compared with the SM prediction \cite{Lenz}
\beq
 \Delta \Gamma_s^{\rm SM} = (0.096 \pm 0.039) \; {\rm ps}^{-1}~.
\label{width_diff_SM}
\eeq
The measurement of $\Delta \Gamma_s$, in principle, constrains NP contributions to $\Gamma^s_{12}$. Note that the present theory predictions are consistent with experimental measurements though it should be kept in mind that theory predictions for $\Delta \Gamma_s$ can contain hadronic uncertainties and constraints on NP from this measurement is not that strong.

In the presence of NP contributions to both   $M^q_{12}$ and  $\Gamma^q_{12}$, the charge asymmetry in Eq.~(\ref{asl}) can be rewritten using Eq.~(\ref{M12def2}) and Eq.~(\ref{Gamdef2}) as
\bea
\label{aslNP}
a^q_{sl} &=& \frac{R^q_{\Gamma}}{R^q_{M}}\frac{ \vert \Gamma^{q,SM}_{12}  \vert}{ \vert M^{q,SM}_{12}  \vert} \sin{(\phi^{q}_{SM}+ \phi^q_{M}-\phi^q_{\Gamma})}.
\eea

\begin{figure}[htb!]
\centering
\includegraphics[width=5.5cm]{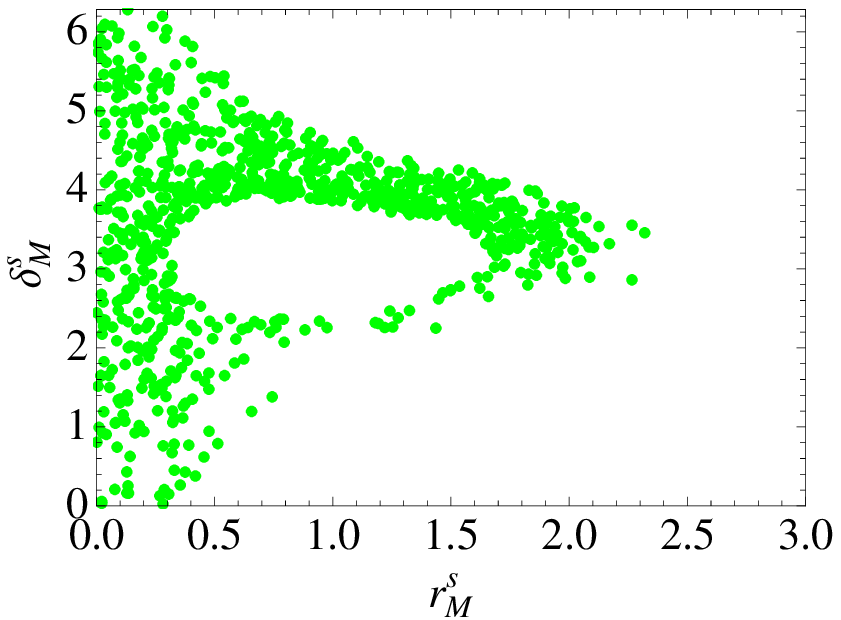} ~\includegraphics[width=5.5cm]{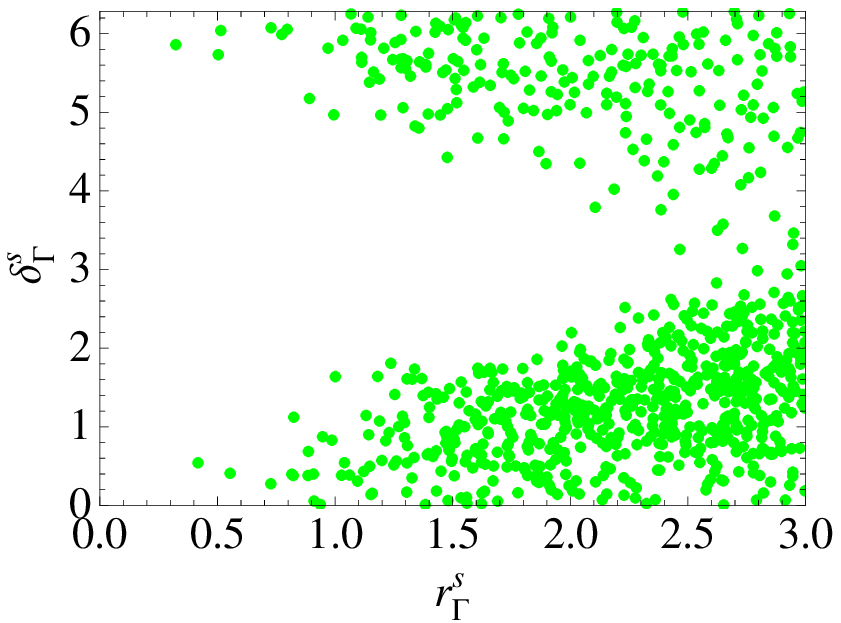}\\
\includegraphics[width=5.5cm]{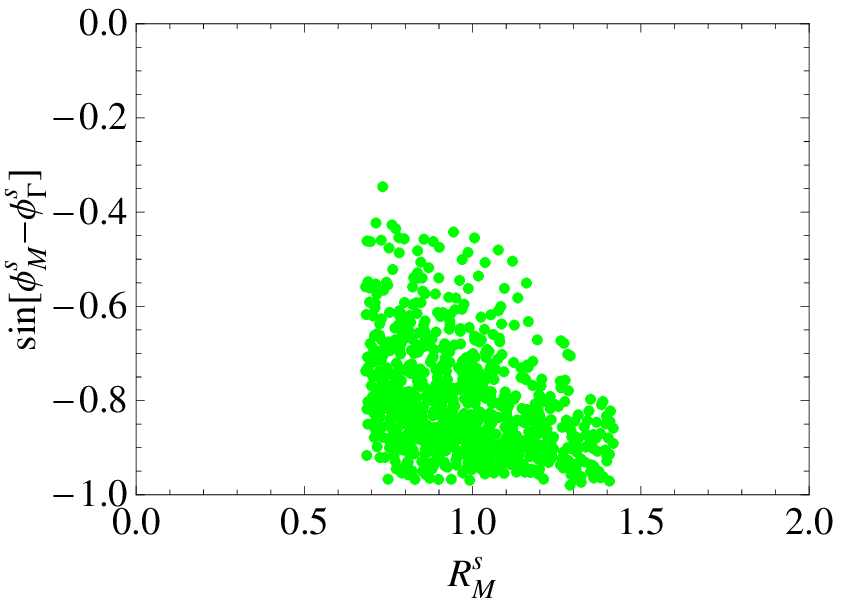} ~\includegraphics[width=5.5cm]{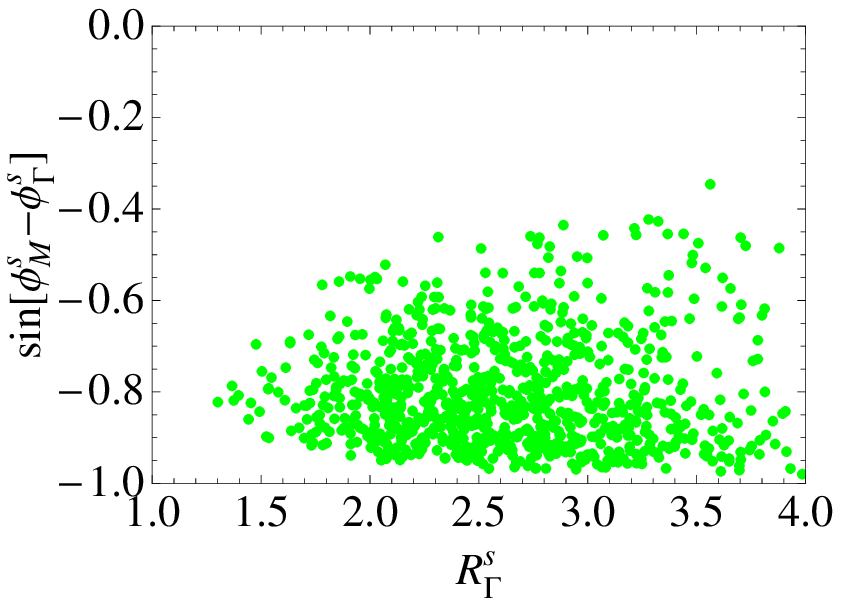}
\caption{ The allowed ranges for  $\delta^s_{M}-r^s_{M}$, $\delta^s_{\Gamma}-r^s_{\Gamma}$, $\sin{(\phi^s_{M}-\phi^s_{\Gamma})}-R^s_{\Gamma}$, and $\sin{(\phi_{M_s}-\phi_{\Gamma_s})}-R_{M_s}$. The scatter points  satisfy the measured $\Delta M_s$, $\Delta \Gamma_s$, and $(a^s_{sl})_{(avg)}$  within the $1 \sigma$ limit.}
 \label{fig.NPobsconfull}
\end{figure}

If one neglects the NP effects to the $\Delta B=1$ effective
Hamiltonian, i.e., $\Gamma^s_{12} = \Gamma_{12}^{s,SM}$ ($R^s_{\Gamma}=0, \phi^s_{\Gamma}=0$), then using Eq.~(\ref{aslNP}) , the charged asymmetry $a^s_{sl}$ is given by %
\be %
a^s_{sl} = \frac{1}{R^s_{M}} \frac{\vert
\Gamma_{12}^{s,SM}\vert}{\vert M_{12}^{s,SM} \vert}
\sin( \phi^s_{M}), %
\ee%
where we have neglected the SM phase $\phi^{s}_{SM} =2 \beta^{SM}_s$. As indicated earlier, using the experimental value for $a^d_{sl}=-0.0047 \pm 0.0046$,
one finds that in order to account for the D\O\ results,
$a^s_{sl}$ must be given by%
\be%
a^s_{sl} = (-14.6 \pm 7.5) \times 10^{-3}. %
\ee%
This implies that %
\be %
\sin \phi^s_{M} = - (2.9 \pm 1.5) \vert R^s_{M} \vert. %
\ee
Using the fact that $\vert R_{M_s} \vert \simeq 1$, one finds that
$\sin \phi_{M_s} \gg 1$, which is unphysical.
Therefore, one concludes that new physics
contribution to $\Gamma^s_{12}$ is necessary to explain the observed CP
charge asymmetry $a^s_{sl}$. This conclusion was also discussed earlier in Refs.~\cite{Dobrescu:2010rh,ew2, D0_theory, lig}. The measured $\Delta M_s$, $\Delta \Gamma_s$, and $(a^s_{sl})_{(avg)}$ in Eqs.~(\ref{MBSexp}), (\ref{width_diff_expt}), and (\ref{asls.avg}), respectively can be used to determine model independently  the ranges for the NP quantities involved in Eq.~(\ref{M12def2}) and ~(\ref{Gamdef2}). In Fig. \ref{fig.NPobsconfull} we show the possible ranges for  $\delta^s_{M}-r^s_{M}$, $\delta^s_{\Gamma}-r^s_{\Gamma}$, $\sin{(\phi^s_{M}-\phi^s_{\Gamma})}-R^s_{\Gamma}$, and $\sin{(\phi_{M_s}-\phi_{\Gamma_s})}-R_{M_s}$. The green scatter points  satisfy the measured $\Delta M_s$, $\Delta \Gamma_s$, and $(a^s_{sl})_{(avg)}$  within the $1 \sigma$ limit. The allowed ranges for  $\delta^s_{M}-r^s_{M}$ and $R^s_{M}$ remain  the same as shown earlier. The phase $\delta^s_{\Gamma}$ is varied in  the range
[0, 2$\pi$], and it is highly constrained below $r^s_{\Gamma} \lsim 0.4 $. Also, one can see that the allowed ranges for  $R^s_{\Gamma} \gsim 1.4$. These results indicate that a  considerably large NP effect  in $\Gamma^s_{12}$  is required to address the observed charge asymmetry $a^s_{sl}$.
We note that similar plots and similar conclusions can be found in recent literature \cite{lig}. 


\section{Effects of New Physics: General case}
\subsection{NP contribution to the decay width $\Gamma^s_{12}$}

In this section we present the SM and NP calculations of the off-diagonal element $\Gamma^s_{12}$ of the  matrix $\Gamma_s$ defined in Eq.~(\ref{Gamdef1}). The results for
$\Gamma^s_{12}$  including general NP are new to the best of our knowledge. We will only consider vector/axial vector operators in the NP Hamiltonian.
\begin{figure}[htb!]
\centering
\includegraphics[width=9.5cm]{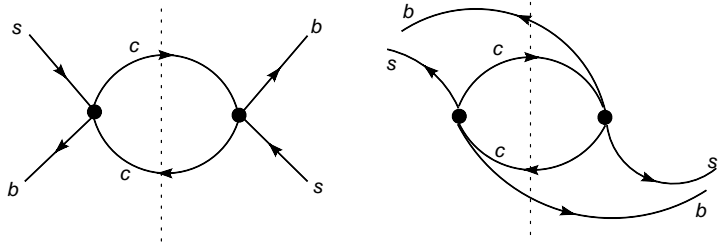}
\caption{A graphical representation of Feynman diagrams contribute
to the leading order $\Gamma^s_{12}$ is shown. }
 \label{fig.width}
\end{figure}
 In the $\bs$ to $\bsbar$ transition amplitude in Eq.~(\ref{Gamdef1}), the dominant contribution comes from
intermediate  $c \bar{c}$ states (see Fig. \ref{fig.width})at the tree level. 
Neglecting CKM-suppressed terms one can write the
effective Hamiltonian for $\Delta B=1$ transitions in the SM  as \cite{Beneke:1998sy}
\bea %
\label{H2}%
{\cal{H}}^{\Delta B=1}_{eff} &=& \frac{G_F}{\sqrt{2}}V^*_{cb}V_{cs} \sum^6_{n=1}\left(
C_n Q_n + h.c \right), %
\eea%
where the operators are
\bea
 \label{eq.op1}
  Q_1 &=& (\bar{b}_i c_j)_{V-A}(\bar{c}_j
s_i)_{V-A},~~~~~~~ Q_2 = (\bar{b}_i c_i)_{V-A}(\bar{c}_j
s_j)_{V-A},\nl Q_3 &=& (\bar{b}_i s_i)_{V-A}  (\bar{q}_j
q_j)_{V-A},~~~~~~~ Q_4 = (\bar{b}_i s_j)_{V-A}  (\bar{q}_j
q_i)_{V-A},\nl Q_5 &=& (\bar{b}_i s_i)_{V-A}  (\bar{q}_j
q_j)_{V+A},~~~~~~~ Q_6 = (\bar{b}_i s_j)_{V-A}  (\bar{q}_j
q_i)_{V+A}. \eea

Here the i and j denote color indices and a summation over these color indices is implied. The notation $(\bar{f_1} f_2)_{V\pm A}$ means $\bar{f_1} \gamma^\mu (1 \pm \gamma_5) f_2$ while
$q=u,d,s,c,b$. $Q_1$, $Q_2$ are color suppressed and color allowed tree-level operators, respectively, and $Q_{3}...Q_{6}$ are QCD penguin operators. The Wilson coefficients of these operators are denoted by $C_{1}... C_{6}$, respectively.
 ~$\Gamma^s_{12}$ is determined up to next-to-leading order  in both $\bar{\Lambda}/m_b$  and $\alpha_s(m_b)$ in \cite{Lenz}, \cite{Beneke:1998sy}, \cite{Beneke:1996gn},  \cite{Dighe:2001sr}, and \cite{ Ciuchini:2003ww}. In the rest of this section, we summarize the leading order  results of $\Gamma^s_{12}$ in the $1/m_b$ expansion and  introduce the effects of new physics to $\Gamma^s_{12}$. We point out that in our numerical analysis, we used the update SM value of $\Delta \Gamma^{SM}_{s}$ in Eq.~(\ref{width_diff_SM}).

In the SM, $\Gamma^s_{12}$ for
the tree-level operators $Q_1$ and $Q_2$ is obtained by computing
the related matrix elements in Eq.~(\ref{eq.trans1}). The
corresponding Feynman diagram is illustrated in
Fig. \ref{fig.width}. $\Gamma^s_{12}$ can be written, to leading order in the $1/m_b$ expansion, as
\cite{Beneke:1998sy}, \cite{Beneke:1996gn}

\bea
\label{Gam12A}
\Gamma^{s,SM}_{12} &=&-\frac{G^2_F m^2_b}{12
\pi(2 M_{B_s})} (V^*_{cb} V_{cs})^2  \Big( G_0(z) <
Q_->+G_{S0}<Q_{S-}>\Big),
 \eea
 where  $z=m^2_c/m^2_b$ and  the
functions $ G_0(z)$ and $G_{S0}$ are given by
\bea
\label{ga}
G_0(z) &=&  \sqrt{1-4 z} \Big((1-z)
C_A+\frac{1}{2}(1-4z)C_B\Big),\nl G_{S0}(z)&=&\sqrt{1-4
z}(1+2z)(C_A-C_B).
 \eea
 The combinations of Wilson coefficients
$C_A$ and $C_B$ are given by
\bea
\label{ca}
C_A &=& \Big(N_C C^2_1+ 2C_1  C_2 \Big), ~~C_B = C^2_2,
 \eea
where $N_C=3$. The two $\Delta B=2$ operators $Q_-$ and $Q_{S-}$ in Eq.~(\ref{Gam12A}) are given by
\bea
 \label{qa} Q_- &=& (\bar{b}_{i }\gamma^\mu(1-\gamma_5)
s_{i})(\bar{b}_{j}\gamma_\mu(1-\gamma_5) s_{j}),~~ Q_{S-} =
(\bar{b}_{i }(1-\gamma_5) s_{i})(\bar{b}_{j}(1-\gamma_5)s_{j}).
\eea
The matrix elements $<Q_->$ and  $<Q_{S-}>$ can be found
using a vacuum insertion method  as \cite{Beneke:1996gn}
 \bea
 \label{qMEa}
 <Q_->&=& \bra{\bar{B}_s} Q_-\ket{B_s}=\frac{8}{3} m^2_{B_s} f^2_{B_s} B_1,\nl
 <Q_{S_-}>&=&\bra{\bar{B}_s} Q_{S_-}\ket{B_s}= -\frac{5}{3} m^2_{B_s} f^2_{B_s} r^s_\chi  B_2,
\eea
where $B_1$ and $B_2$ are bag parameters, and the quantity $r^s_\chi$ is defined by
\bea
 \label{rchi} r^s_\chi &=&
\frac{m^2_{B_s}}{(\bar{m}_b(m_b)+\bar{m}_s(m_b))^2}.
 \eea

The contributions of the all six operators $Q_i (i=1..6)$ in Eq.~(\ref{eq.op1}) to
$\Gamma^s_{12}$ are given by
\bea
\label{Gam12B}
\Gamma^{s,SM}_{12} &=&-\frac{G^2_F
m^2_b}{12 \pi(2 M_{B_s})} (V^*_{cb} V_{cs})^2   \Big(
G^\prime_0(z)< Q_->+ G^\prime_{S0}(z)<Q_{S-}>\Big),
\eea
where the
functions $G^\prime_0(z)$ and $G^\prime_{S0}(z)$ are given by
\cite{Beneke:1996gn}
\bea
\label{gB} G^\prime_0(z) &=&  \sqrt{1-4
z} \Big ((1-z) C^\prime_A+\frac{1}{2}(1-4z)C^\prime_B +3 z
C^\prime_C \Big),\nl G^\prime_{S0}(z)&=&\sqrt{1-4 z}(1+2z)(
C^\prime_A-C^\prime_B). \eea The combinations of Wilson
coefficients $ C^\prime_A$ and $ C^\prime_B$ are given by \bea
\label{cp}
 C^\prime_A &=& \Big(N_C(C_1+C_3)^2+2(C_1+C_3)(C_2+C_4)+2 C_5 C_6+N_C C^2_5 \Big),\nl
 C^\prime_B &=& \Big((C_2+C_4)^2+C^2_6 \Big),\nl
 C^\prime_C &=& 2\Big(N_C(C_1+C_3)C_5   +(C_1+C_3)C_6 +(C_2+C_4)C_5  +(C_2+C_4)C_6 \Big).
\eea
 Setting $C_{i=3..6} = 0$ in Eq.~(\ref{Gam12B}) one recovers  Eq.~(\ref{Gam12A}). The effects of $C_{i=3..6}$ on $\Gamma^{s,SM}_{21}$ are small.

\par Next we consider the contributions to $\Gamma^s_{12}$ from  NP operators involving  $b \to c \bar{c} s$ transitions. Note that these operators can also contribute to the lifetimes of the $b$ hadrons and we will
consider constraints on these operators from lifetime measurements in our numerical analysis.
In general NP can also contribute to $ b \to q \bar{q} s$ transitions with $q=u,d,s$. However, we note that in the SM these transitions are
suppressed and measurements in decays like $B \to K \pi$, $B \to \phi K$ etc. already constrain these NP contributions. In other words, an NP contribution to $ b \to q \bar{q} s$ with $q=u,d,s$ cannot significantly affect $\Gamma_{12}^s$ and hence we ignore these contributions.
The $b \to c \bar{c} s$ transitions, on the other hand, are tree level in the SM and measurements of decays with the underlying
$b \to c \bar{c} s$ transitions, such as $B_d \to D_s D, J/\psi K_s$ etc.
might still allow for NP in $b \to c \bar{c} s$ transitions that are not small.
In this section, we  only present model independent results, and for the numerics in the following section we consider the RS model with split fermions. The RS model with split fermions can also generate
NP contributions to $ b \to s \tau^+ \tau^-$ which are not that well constrained from experiments. However, these transitions are generated by
the exchange of KK electroweak bosons and hence they are much smaller than
the $b \to c \bar{c} s$ transitions that are generated by KK gluon exchange.

The $\Delta B=1$ effective weak Hamiltonian for NP
operators is written as
\bea
 \label{HNP}
 {\cal{H}}^{NP}_{eff}
&=& \Big(\lambda_{LL} Q_{LL}+ \lambda^\prime_{LL} Q^\prime_{LL}+
\lambda_{LR} Q_{LR}+ \lambda^\prime_{LR} Q^\prime_{LR}\nl &&
+\lambda_{RR} Q_{RR}+\lambda^\prime_{RR}Q^\prime_{RR}+\lambda_{RL}
Q_{RL}+\lambda^\prime_{RL}Q^\prime_{RL}\Big),
 \eea
where
\bea \label{qNP} Q_{LL} &=&  (\bar{b}_i s_i)_{V-A}  (\bar{c}_j
c_j)_{V-A},~~~~~~~
 Q^\prime_{LL} = (\bar{b}_i s_j)_{V-A}  (\bar{c}_j c_i)_{V-A},\nl
Q_{LR} &=&  (\bar{b}_i s_i)_{V-A}  (\bar{c}_j c_j)_{V+A},~~~~~~~
 Q^\prime_{LR} =(\bar{b}_i s_j)_{V-A}  (\bar{c}_j c_i)_{V+A},\nl
Q_{RR} &=& (\bar{b}_i s_i)_{V+A}  (\bar{c}_j c_j)_{V+A},~~~~~~~
Q^\prime_{RR}=  (\bar{b}_i s_j)_{V+A}  (\bar{c}_j c_i)_{V+A},\nl
Q_{RL} &=& (\bar{b}_i s_i)_{V+A}  (\bar{c}_j c_j)_{V-A}, ~~~~~~~
Q^\prime_{RL} =  (\bar{b}_i s_j)_{V+A}  (\bar{c}_j c_i)_{V-A}.
\eea

 The eight couplings $\lambda_{AB}$, and $ \lambda^\prime_{AB}$ (A, B = L, R) are in general complex and
can be determined from  specific models. Thus, the total effective Hamiltonian can be written as
\bea
\label{HTot} {\cal{H}}_{eff} &=& {\cal{H}}^{SM}_{eff}
+{\cal{H}}^{NP}_{eff},\nl
 {\cal{H}}_{eff} &=&
\frac{G_F}{\sqrt{2}}(V^*_{cb}V_{cs}) \Big((C_1+C_3 + C^{LL}_3)
Q_1+(C_2+C_4+  C^{LL}_4) Q_2+ (C_5+  C^{LR}_5) Q_5+ (C_6+  C^{LR}_6) Q_6
\nl &&  + C^{RR}_3 Q_{RR}+ C^{RR}_4 Q^\prime_{RR}+
C^{RL}_5 Q_{RL}+C^{RL}_6 Q^\prime_{RL}\Big),
 \eea
where the new Wilson coefficients (A, B = L, R) are
\bea
 \label{CTot}
  C^{AA}_3 &=& \frac{\sqrt{2} \lambda_{AA}}{G_F V^*_{cb}V_{cs}}, ~~~~ C^{AA}_4 =\frac{\sqrt{2} \lambda^\prime_{AA}}{G_F V^*_{cb}V_{cs}}, \nl
  C^{AB}_5 &=& \frac{\sqrt{2} \lambda_{AB}}{G_F V^*_{cb}V_{cs}}, ~~~~ C^{AB}_6 =\frac{\sqrt{2} \lambda^\prime_{AB}}{G_F V^*_{cb}V_{cs}}~~~(A\neq B).
  \eea
The NP contributions to $\Gamma^s_{21}$ are obtained by  computing the related matrix elements in
Eq.~(\ref{eq.trans1}) to the leading order in the $1/m_b$ expansion using the $\overline{MS}$ scheme. The NP contributions contain both pure  NP and SM-NP interference terms. In general  the latter dominate NP contributions due to the large SM Wilson coefficients. We obtain NP contributions to  $\Gamma^s_{21}$ as
\bea
\label{GammaNPtotal}
\Gamma^{s,NP}_{12 } &=& \Gamma^{s,LL}_{12 }+\Gamma^{s,RR}_{12 }+\Gamma^{s,mix}_{12 },
\eea
where   $\Gamma^{s,LL}_{12 }$ , $\Gamma^{s,RR}_{12
 }$, and $\Gamma^{s,mix}_{12 }$  contain the contributions from  LL and LR, RR and RL, and all possible  type of operators, respectively. They can be expressed in terms of
the matrix elements of eight $\Delta B=2 $ operators
\bea
\label{newB2op}
 Q_- &=& (\bar{b}_{i }s_{i})_{V-A}(\bar{b}_{j} s_{j})_{V-A},~~~~~~~ Q_{S_-} =
(\bar{b}_{i } s_{i})_{S-P}(\bar{b}_{j}s_{j})_{S-P},\nl
Q_+ &=& (\bar{b}_{i }s_{i})_{V+A}(\bar{b}_{j} s_{j})_{V+A},~~~~~~~ Q_{S_+} =
(\bar{b}_{i } s_{i})_{S+P}(\bar{b}_{j}s_{j})_{S+P},\nl
Q_\mp &=& (\bar{b}_{i }s_{i})_{V-A}(\bar{b}_{j} s_{j})_{V+A},~~~~~~~ Q_{S_\mp}
= (\bar{b}_{i }s_{i})_{S-P}(\bar{b}_{j}s_{j})_{S+P},\nl
Q_\pm &=& (\bar{b}_{i}s_{i})_{V+A} (\bar{b}_{j}s_{j})_{V-A},~~~~~~~  Q_{S_\pm} =  (\bar{b}_{i}s_{i})_{S+P} (\bar{b}_{j}  s_{j})_{S-P},
\eea
where S$\pm$ P = 1 $\pm \gamma_5$. The explicit forms of $\Gamma^{s,LL}_{12 }$, $\Gamma^{s,RR}_{12
 }$, and $\Gamma^{s,mix}_{12 }$  are
\bea
\label{Gam12C}
\Gamma^{s,LL}_{12 } &=&-\frac{G^2_F m^2_b}{12 \pi(2 M_{B_s})} (V^*_{cb} V_{cs})^2
\Big( G^{\prime \prime} _0(z)< Q_->+ G^{\prime \prime}_{S0}(z)<Q_{S-}>\Big),\nl
\Gamma^{s,RR}_{12 }&=&-\frac{G^2_F m^2_b}{12 \pi(2 M_{B_s})} (V^*_{cb} V_{cs})^2
\Big( \hat{G} _0(z)< Q_+>+ \hat{G}_{S0}(z)<Q_{S+}>\Big),\nl
\Gamma^{s,mix}_{12 }&=&-\frac{G^2_F m^2_b}{12 \pi(2 M_{B_s})}
(V^*_{cb} V_{cs})^2 \Big(\bar{G}_1(z)+ \bar{G}_2(z)+\bar{G}_3(z)+\bar{G}_4(z)\Big)
, \eea
where the G functions  are
\bea
\label{GPP}
 G^{\prime \prime}_0(z) &=& \sqrt{1-4 z} \Big ((1-z)C^{\prime \prime}_A+\frac{1}{2}(1-4z)C^{\prime \prime}_B +3 z
C^{\prime \prime}_C \Big),\nl
G^{\prime\prime}_{S0}(z)&=&\sqrt{1-4
z}(1+2z)( C^{\prime\prime}_A-C^{\prime\prime}_B),\nl
\hat{G}_0(z) &=& \sqrt{1-4 z} \Big ((1-z) \hat{C}_A+\frac{1}{2}(1-4z)\hat{C}_B
+3 z \hat{C}_C \Big),\nl
\hat{G}_{S0}(z)&=&\sqrt{1-4z}(1+2z)(\hat{C}_A-\hat{C}_B),\nl
 \bar{G}_1(z) &=& 2\sqrt{1-4z}(C_1+C_3+C^{LL}_3) \Big[\Big((1-z)(N_C C^{RL}_5+ C^{RL}_6 ) +3
z (N_C C^{RR}_3+ C^{RR}_4)\Big)<Q_\mp> \nl && +(1+2z)
(N_C C^{RL}_5+ C^{RL}_6 ) <Q_{S\mp}>\Big],\nl
\bar{G}_2(z)&=& 2\sqrt{1-4 z}(C_2+C_4+C^{LL}_4 ) \Big[ \Big(3 z
C^{RR}_3+(1-z)C^{RL}_5 \Big) <Q_\mp> + (1+2z)C^{RL}_5
<Q_{S_\mp}>\nl && -\frac{1}{2}(1+2z) C^{RL}_6 <Q_{\pm}>-2\Big(3
z C^{RR}_4 +(1-z)C^{RL}_6 \Big)<Q_{S_\pm}> \Big],\nl
\bar{G}_3(z)&=& 2\sqrt{1-4 z}(C_5+C^{LR}_5 )\Big[\Big((1-z)(N_C
C^{RR}_3+C^{RR}_4 )+3 z ( N_C C^{RL}_5+C^{RL}_6 )\Big)
<Q_\mp>\nl && +(1+2z) \Big(N_C C^{RR}_3+C^{RR}_4
\Big)<Q_{S_\mp}>\Big],\nl
\bar{G}_4(z)&=& 2\sqrt{1-4 z}(C_6+C^{LR}_6 ) \Big[\Big((1-z)C^{RR}_3+3 z C^{RL}_6 \Big)<Q_\mp> \nl &&
+(1+2z)C^{RR}_3 <Q_{S_\mp}> +3 z(- C^{RR}_4   <Q_\pm>+2 C^{RL}_6  < Q_{S_\pm}>)\Big].
\eea
 The combinations of Wilson coefficients are
 \bea
 \label{CPP}
 C^{\prime}_A &=& \Big[N_C\Big((C^{LL}_3)^2+(C^{LR}_5)^2\Big) + 2 C^{LL}_3 C^{LL}_4+2 C^{LR}_5 C^{LR}_6 + 2 C^{LL}_3 ( N_C (C_1 + C_3)+C_2 + C_4 )\nl && + 2  C^{LL}_4 ( C_1 +C_3) +2 C^{LR}_5 (N_C C_5 + C_6 ) + 2 C^{LR}_6 C_5 \Big],\nl
 C^{\prime}_B &=& \Big[(C^{LL}_4)^2+(C^{LR}_6)^2 +2 C^{LL}_4 (C_2 + C_4) + 2 C^{LR}_6 C_6 )\Big], \nl
 C^{\prime}_C &=& \Big[2 C^{LL}_3 (N_C  C^{LR}_5 + C^{LR}_6) +  2  C^{LL}_4 (C^{LR}_5 + C^{LR}_6) + 2 C^{LL}_3 (N_C  C^{LR}_5 + C^{LR}_6) \nl && + 2 C^{LL}_4 (C_5 + C_6) +
    2  C^{LR}_5 ( N_C (C_1 + C_3)+C_2 + C_4 ) + 2  C^{LR}_6 (C_1 + C_2 + C_3 + C_4)\Big], \nl
\tilde{C}_A &=&
\Big[N_C\Big((C^{RR}_3)^2+(C^{RL}_5)^2\Big)+2 C^{RR}_3 C^{RR}_4\Big],\nl
\tilde{C}_B &=&\Big[(C^{RR}_4)^2+(C^{RL}_6)^2 \Big],\nl
 \tilde{C}_C &=& \Big[ 2(N_C C^{RR}_3 C^{RL}_5   + C^{RR}_3 C^{RL}_6 +C^{RR}_4 C^{RL}_5  +C^{RR}_4 C^{RL}_6)\Big].
  \eea
%

The matrix elements $<Q>=\bra{\bar{B}_s} Q\ket{B_s}$ of the
operators in Eq.~(\ref{newB2op}) are given by \cite{Buras:20001ra}
\bea
\label{Bagdef}
 <Q_->&=&<Q_+>=\frac{8}{3}m^2_{B_s} f^2_{B_s}  B_1,\nl
 <Q_{S_-}> &=&<Q_{S_+}>=  -\frac{5}{3} m^2_{B_s} f^2_{B_s} r^s_\chi  B_2,\nl
 <Q_{S\mp}> &=&  <Q_{S\pm}>= 2 m^2_{B_s} f^2_{B_s} r^s_\chi  B_4,\nl
 <Q_\mp>&=&<Q_\pm> =-\frac{4}{3} m^2_{B_s} f^2_{B_s} r^s_\chi  B_5,
 \eea
 where $B_i$'s are the bag parameters and their numerical values for $B^0_s-\bar{B}^0_s$ system in the $\overline{MS}$-NDR scheme at $m_b= 4.6$ Gev can be found in \cite{Becirevic:2001xt}.
\subsection{NP contribution to the mass parameter $M^s_{12}$}

In this section we present the SM and NP calculations of the off-diagonal element $M^s_{12}$ of the  matrix $M_q$ defined in Eq.~(\ref{M12def}). The effective Hamiltonian for $\Delta B=2$ transition that generates $B^0_s$-$\bar{B}^0_s$ mixing in the SM  can be written as \cite{K2}
\bea
\label{delB2SM}
{\cal{H}}^{\Delta B=2, SM}_{eff} &=& \frac{G^2_F}{16 \pi^2} M^2_{W} (V_{tb}V^*_{ts})^2  \eta_{B_s} S_0(x_t) (\bar{b} s)_{V-A}(\bar{b} s)_{V-A},
\eea
where $\eta_{B_s}\simeq 0.551$ \cite{Laiho:2009eu} is the QCD correction, and the loop function  $S_0(x_t)$ ($x_t=m^2_t/M^2_W$) is given by \cite{Buras:1990fn}
\bea
\label{S0}
S_0(x_t) &=& \frac{4 x_t-11 x^2_t+x^3_t}{4(1-x_t)^2}-\frac{3 x^3_t ln(x_t)}{2(1-x_t)^3}.
\eea
The SM contribution to $M^s_{12}$ can be  obtained using  Eq.~(\ref{M12def}) as
\bea
\label{delM12SM}
M^{s,SM}_{12}&=& \frac{G^2_F}{12 \pi^2} M^2_{W} (V_{tb}V^*_{ts})^2  \eta_{B_s} S_0(x_t) m_{B_s} f^2_{B_s} B_1 .
\eea

Next we consider  NP effects in the $\Delta B=2 $ transition.
The effective Hamiltonian for this transition is written as
\bea
 \label{HNPdelB2}
 {\cal{H}}^{\Delta B=2,NP}_{eff}
&=& \Big(\delta_{LL} Q_{-}+ \delta^\prime_{LL} Q^\prime_{-}+
\delta_{LR} Q_{\mp}+ \delta^\prime_{LR} Q^\prime_{\mp}\nl &&
+\delta_{RR} Q_{+}+\delta^\prime_{RR}Q^\prime_{+}+\delta_{RL}
Q_{\pm}+\delta^\prime_{RL}Q^\prime_{\pm}\Big),
 \eea
where $\delta_{AB}$ (A, B = L, R) are  NP couplings. The
 operators $Q_{-}$, $Q_{+}$, $Q_{\mp}$, and $Q_{\pm}$ are given in Eq. (\ref{newB2op}) and their  color suppressed counterpart operators  can be written as
\bea
\label{newB2opcolorsup}
 Q^\prime_- &=& (\bar{b}_{i }s_{j})_{V-A}(\bar{b}_{j} s_{i})_{V-A},~~~~~~~
Q^\prime_+ = (\bar{b}_{i }s_{j})_{V+A}(\bar{b}_{j} s_{i})_{V+A},\nl
Q^\prime_\mp &=& (\bar{b}_{i }s_{j})_{V-A}(\bar{b}_{j} s_{i})_{V+A},~~~~~~~
Q^\prime_\pm = (\bar{b}_{i}s_{j})_{V+A} (\bar{b}_{j} s_{i})_{V-A}.
\eea
 Using  Eq.~(\ref{M12def}) and applying the Fierz transformation, one can obtain the contribution of these NP operators to $M^s_{12}$ as
\bea
\label{M12sNP}
M^s_{12,NP} &=& \frac{1}{2 m_{B_s}} \Big( (\delta^\prime_{LL} + \delta_{LL})<Q_-> + (\delta^\prime_{RR}+\delta_{RR} )<Q_-> + (\delta_{LR}+  \delta_{RL}) <Q_\mp>\nl &&  -2(\delta^\prime_{LR} +\delta^\prime_{RL}) <Q_{S_\mp}>\Big).
\eea
For certain classes of models including the RS model with split fermions the following relations hold:
\bea
\delta_{LL}&=& -1/3 \delta^\prime_{LL}, ~~\delta_{RR}= -1/3 \delta^\prime_{RR},~~
\delta_{LR}=-1/3  \delta ^\prime_{LR},~~\delta_{RL}= -1/3  \delta ^\prime_{RL},~~
\delta^\prime_{LR}=\delta^\prime_{RL}.
\eea
One can then obtain
\bea
\label{M12NP}
M^s_{12,NP} &=& \frac{4}{3} m_{B_s} f^2_{B_s} \Big[ \frac{2}{3} \Big(\delta^\prime_{LL}+\delta^\prime_{RR} \Big) B_1  +\delta^\prime_{LR} \Big(\frac{1}{3}   r^s_\chi B_5 - 3 r^s_\chi B_4\Big) \Big],
\eea
 where the bag parameters are defined  in Eq.~(\ref{Bagdef}).

\section{New source of flavor and CP violation in the RS model}
The warped extra dimension model has been proposed as a solution of the
hierarchy problem. In the original RS model, the SM fields are
localized to one of the boundaries and gravity is allowed to
propagate in the bulk. However, it was realized that  scenarios
with SM gauge bosons and fermions in the bulk may lead to a new
 geometrical interpretation for the hierarchy
of quark and lepton masses. The Higgs field has to be confined to
the TeV brane in order to obtain the observable masses of the $W$
and $Z$ gauge bosons.

We will consider the scenario of Ref.\cite{RS}, based on the
metric %
\be %
ds^2 = e^{-2\sigma(y)} \eta_{\mu \nu} dx^{\mu} dx^{\nu} + d y^2, %
\ee%
where $\sigma(y)=\kappa \vert y \vert$ and $\kappa
\sim M_P$ is the curvature scale determined by the negative
cosmological constant in the five dimensional bulk. The fermion
fields reside in the bulk of this nonfactorizable geometry and can be
decomposed as %
\be %
\Psi(x,y) = \frac{1}{\sqrt{2 \pi r_c}}
\sum_{n=0}^{\infty} \psi^{(n)}(x) e^{2 \sigma(y)} f^{(n)}(y).%
\ee%
Here $r_c$ is the radius of the compactified fifth dimension on
an orbifold $S_1/Z_2$ so that the bulk is a slice of $AdS_5$ space
between two four dimensional boundaries. The left-handed zero mode wave
function is given by \cite{Gherghetta:2000qt} %
\be %
\label{zeromodeKK}
f^{(0)}_L(c_{f_\alpha},y)=\frac{e^{-c_{f_\alpha} \sigma(y)}}{N_0}. %
\ee%
where $c_{f_\alpha}=m_{f_\alpha}/\kappa$ and $m_{f_\alpha}$ is the bulk mass term.
Using the orthonormal condition:
\bea
\frac{1}{2\pi r_c} \int_{-\pi r_c}^{\pi r_c} dy e^{\sigma(y)} f^{(0)}_L(c_{f_\alpha},y) f^{(0)}_L(c_{f_\alpha},y) =1,
\eea
 one finds that $N_0$
is given by %
\be %
N_0 = \sqrt{\Frac{e^{\pi \kappa r_c (1-2 c_{f_\alpha})} -1}{\pi
\kappa r_c(1- 2 c_{f_\alpha})}}. %
\ee %
The right-handed zero mode wave function can be obtained from %
\be%
f^{(0)}_R(c_{f_\alpha},y) = f^{(0)}_L(-c_{f_\alpha},y). %
\ee %
The tower of fermion KK excited states is not relevant to our
discussion here. Note that fermions with $c_f > 1/2$ are localized
near the Planck brane at $y=0$ and fermions with $c_f < 1/2$ are
localized near the TeV brane at $y=\pi r_c$.

The massless gauge fields that propagate in this curved background
can be decomposed as
\cite{Gherghetta:2000qt} %
\be %
A_{\mu}(x,y) = \frac{1}{\sqrt{2 \pi
r_c}} \sum_{n=0}^{\infty} A_{\mu}^{(n)}(x) f^{(n)}_A(y), %
\ee %
with $f^{(n)}_A(y)$ is dimensionless. The  n-th ($n > 0$) mode function is given as %
\be%
f^{(n)}_A(y) = \frac{e^{\sigma(y)}}{N_n} \left[
J_1\left(\frac{m^{(n)}_A}{\kappa} e^{\sigma(y)} \right) + b_A(m^{(n)}_A)
Y_1\left( \frac{m^{(n)}_A}{\kappa} e^{\sigma(y)}\right) \right], %
\ee%
where $J_1$ and $Y_1$ are the J- and Y-type Bessel functions of order one and
\bea
 b_A(m^{(n)}_A)=- \frac{J_0(m^{(n)}_A/\kappa)}{Y_0(m^{(n)}_A/\kappa)}.
 \eea
 The coupling of the gauge KK modes to  the fermion at the vertex $q^{\alpha}_{L(R)} \bar{q}^\beta_{L(R)} g^{(n)}$ is given by %
\be %
g^{(n)}(c_{f_\alpha})_{L(R)} = \frac{g^{(5)}}{(2\pi r_c)^{3/2}} \int_{-\pi r_c}^{\pi r_c}
e^{\sigma(y)} f^{(0)}_{L(R)}(c_{f_\alpha},y) f^{(0)}_{L(R)}(c_{f_\alpha},y) f^{(n)}_A(y) dy . %
\ee%
The nonuniversal parameters $c_{f_\alpha}$ lead to nonuniversal couplings
to the KK state of the gluon. In the basis of mass eigenstates we
have the following flavor dependent couplings: %
\be %
\label{RScoup}
(U^{u,d(n)}_{L(R)})_{\alpha \beta} =( V^{u,d^+}_{L(R)})_{\alpha \gamma}~ g^{(n)}_{L(R)}(c_{f_\gamma})~ (V^{u,d}_{L(R)})_{\gamma \beta}, %
\ee%
where the $g^{(1)}_{L(R)}$ is given by %
\be %
\label{gsKK} %
g^{(1)}_{L(R)}(c_{f_\alpha}) = g \left(\frac{1-2
c_{f_\alpha}}{e^{\pi \kappa r_c(1-2c_{f_\alpha})} -1}\right)
\frac{\kappa}{N_0} \int_0^{\pi r_c} e^{(1-2c_{f_\alpha})\kappa y}
\left[J_1\left(\frac{m^{(1)}_A}{\kappa} e^{\kappa y}\right) +
b_A(m^{(1)}_A)
Y_1\left(\frac{m^{(1)}_A}{\kappa} e^{\kappa y}\right)\right]. %
\ee%
The tree-level relation between the 5D and 4D QCD couplings $g_5$ and g is $g=g_5/\sqrt{2 \pi r_c}$. However, at the one loop level the relation between these two couplings also depends on the value of the brane-localized kinetic terms for the bulk gauge fields, and there can be significant corrections to the tree level relation. A detailed discussion on this topic can be found in Ref.~\cite{Agashe:2008uz}. Finally, the unitary
matrices $V_{L(R)}^{u,d}$ diagonalize the up/down quark mass matrix $M^{u,d}_{\alpha \beta}=(v/\sqrt{2}) Y^{u,d}_{\alpha \beta}$,
which
is given in this model as %
\be %
Y^{u,d}_{\alpha \beta} = \frac{l_{\alpha \beta}}{\pi \kappa r_c} f^{(0)}_L (c_{Q},y=\pi r_c)f^{(0)}_R (c_{u,d},y=\pi r_c). %
\ee%

The dimensionless parameters $l_{\alpha \beta}$ are defined as
\bea
l_{\alpha \beta}=\lambda_{\alpha \beta}^{(5)} \sqrt{\kappa},
 \eea
 where $\lambda_{\alpha \beta}^{(5)}$ are
the $5D$ Yukawa couplings which are free parameters to be fixed by
the observable masses and mixing. In Table \ref{tabinp}, we
present an example of the $c_f$ parameters that leads to the
correct quark masses and mixing with $\lambda^{(5)} \sim {\cal
O}(1)$. The corresponding first-KK gluon coupling constants are
given
by%
\bea %
g^{(1)}_L(c_{Q_1}) &=&-.199,~~~~~~~~~~
g^{(1)}_L(c_{Q_2})=-.198,~~~~~~~~~~ g^{(1)}_L(c_{Q_3}) = 1.496,\nl
g^{(1)}_R(c_{D_1}) &=& -.191,~~~~~~~~~~
g^{(1)}_L(c_{D_2})=-.191,~~~~~~~~~~g^{(1)}_L(c_{D_3}) = -0.198,\nl
g^{(1)}_R(c_{U_1}) &=& -.199,~~~~~~~~~~
g^{(1)}_L(c_{U_2})=-.195,~~~~~~~~~~g^{(1)}_L(c_{U_3}) = 3.38.
\eea %
We note that the value of $c_{D3}$ in  Table. \ref{tabinp}  is larger than for the 1st/2nd
generation. Such a choice of c's is not consistent with anarchic
5D Yukawa couplings discussed in Ref~\cite{ag}. We also point out that
  for simplicity, we have used here the tree-level relation between the 5D and 4D QCD couplings. As can be seen from these values, $g^{(1)}$ is of order one.
In fact, this general conclusion  can be obtained with any
other values of $c_f$. One can easily show that for $c_f > 1/2$
the coupling $g^{(1)}$ approaches zero, while for $c_f < 1/2$ one
finds $g^{(1)} \lsim 4$. We expect the general features about $g^{(1)}$ to remain true even when the one loop matching of the 5D and 4D QCD couplings is used.
\begin{figure}[t]
\begin{center}
\epsfig{file=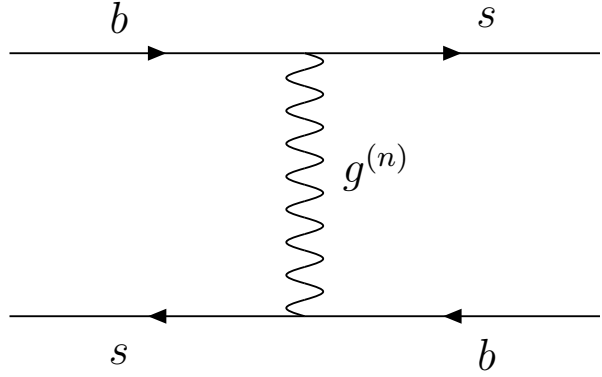, width=8cm,height=5cm,angle=0}
\end{center}
\caption{The gluon KK contribution to $ B^0_s-\bar{B}^0_s $ mixing is shown. }
\label{Bs}
\end{figure}

In the RS model, the effective Hamiltonian for $\Delta B= 2$
transition can be generated at tree-level via the exchange of a  KK gluon, as shown in Fig. \ref{Bs}.
The $\Delta B=2$ effective Hamiltonian is given by \cite{SM},
\bea
\label{delB2RSModel}
{\cal{H}}^{\Delta B=2,KK}_{eff} &=&\sum_{A,B} \frac{1}{4 m^2_{KK}} (U^{d(1)}_A)_{32} (U^{d(1)}_B)_{32} \Big[(\bar{b}_i \gamma^\mu P_A s_{j})(\bar{b}_{j} \gamma_\mu P_B s_{i})-\frac{1}{3} (\bar{b}_i \gamma^\mu P_A s_{i})(\bar{b}_{j} \gamma_\mu P_B s_{j})\Big],
\eea
with  A, B = L, R, and $P_{L(R)} = 1/2(1\mp \gamma_5)$. The contribution to $M^s_{12}$ in the RS model can be obtained by comparing Eq.~(\ref{delB2RSModel}) with Eq.~(\ref{HNPdelB2}), and the couplings can be expressed as,
\bea
\label{delKKcoup}
\delta^\prime_{LL} &=&  \frac{1}{16 m^2_{KK}} (U^{d(1)}_L)_{32} (U^{d(1)}_L)_{32}, ~~~\delta^\prime_{RR} =  \frac{1}{16 m^2_{KK}} (U^{d(1)}_R)_{32} (U^{d(1)}_R)_{32} ,\nl
\delta^\prime_{LR} &=&  \frac{1}{16 m^2_{KK}} (U^{d(1)}_L)_{32} (U^{d(1)}_R)_{32}, ~~~\delta^\prime_{RL} =  \frac{1}{16 m^2_{KK}} (U^{d(1)}_R)_{32} (U^{d(1)}_L)_{32}.
\eea
Thus, Eq.~(\ref{M12NP}) reduces to
\bea
\label{MsKeqK}
M^{s,KK}_{12} &=& \frac{1}{12 m^2_{KK} } m_{B_s} f^2_{B_s} \Big[ \frac{2}{3} \Big((U^{d(1)}_L)^2_{32}+(U^{d(1)}_R)^2_{32} \Big) B_1  + (U^{d(1)}_L)_{32} (U^{d(1)}_R)_{32} \Big(\frac{1}{3}   r^s_\chi B_5 - 3 r^s_\chi B_4\Big)  \Big].
\eea

This result agrees with those in ~\cite{K2}. If $M^{s,KK}_{12}$
is dominated by the first term, then the ratio $r_{M_s}$ in
Eq. ~(\ref{M12def3}) can be obtained  for the RS model as %
\bea%
\label{rmKK} %
r^{KK}_{M_s} &=& \Big\vert\frac{M^{s,KK}_{12}}{M^{s,SM}_{12}}
\Big\vert =\frac{2 \pi^2}{3 m^2_{KK} G^2_F
M^2_W}\frac{(U^{d(1)}_L)^2_{32}}{ (V_{tb}V^*_{ts})^2  \eta_{B_s}
S_0(x_t) }\sim\Big(\frac{2415}{m_{KK}}\Big)^2
\Big(\frac{(U^{d(1)}_L)_{32}}{ V_{tb}V^*_{ts}}\Big)^2. %
\eea %
Therefore, $r^{KK}_{M_s} \sim$ $\cal{O}$(1) requires
 $m_{KK} \sim $ 2.4 TeV if we assume
$(U^{d(1)}_L)_{32}\sim V_{tb}V^*_{ts}$.
 If $(U^{d(1)})_{32} \simeq {\cal O}(1)$, the
experimental limits on $\Delta M_{B_s}$ implies that $m_{KK} \gsim
10 {\rm TeV}$, which imposes stringent constraint on the associated
compactification scale.


The $\Delta B=1$ effective Hamiltonian for the  $b\rightarrow s \bar{c} c$ transition  in the RS Model can be written as
\bea
\label{delB1RSmodel}
{\cal{H}}^{\Delta B=1,KK}_{eff} &=& \frac{1}{4 m^2_{KK}} (U^{d(1)}_A)_{32} (U^{u(1)}_B)_{22} \Big[(\bar{b}_i \gamma^\mu P_A s_{j})(\bar{c}_{j} \gamma_\mu P_B c_{i})-\frac{1}{3} (\bar{b}_i \gamma^\mu P_A s_{i})(\bar{c}_{j} \gamma_\mu P_B c_{j})\Big].
\eea

The contribution of the RS model to  $\Gamma^s_{12}$ can be obtained
using Eq.~(\ref{GammaNPtotal}) with the couplings \bea
\label{RScoupGamma} \delta^\prime_{LL} &=&  \frac{1}{16 m^2_{KK}}
(U^{d(1)}_L)_{32} (U^{u(1)}_L)_{22}, ~~
\delta_{LL}=-\frac{1}{3}\delta^\prime_{LL},\nl \delta^\prime_{LR}
&=&  \frac{1}{16 m^2_{KK}} (U^{d(1)}_L)_{32} (U^{u(1)}_R)_{22}, ~~
\delta_{LR}=-\frac{1}{3}\delta^\prime_{LR},\nl \delta^\prime_{RR}
&=&  \frac{1}{16 m^2_{KK}} (U^{d(1)}_R)_{32} (U^{u(1)}_R)_{22}, ~~
\delta_{RR}=-\frac{1}{3}\delta^\prime_{RR},\nl \delta^\prime_{RL}
&=&  \frac{1}{16 m^2_{KK}} (U^{d(1)}_R)_{32} (U^{u(1)}_L)_{22}, ~~
\delta_{RL}=-\frac{1}{3}\delta^\prime_{RL}. \eea The corresponding
Wilson coefficients in the RS Model can be obtained from
Eq. (\ref{CTot}).

\section{Numerical Results}
The numerical inputs for the parameters in the SM  \cite{pdg}
and RS model are summarized  in Table.~\ref{tabinp}. The values of the bag parameters in the
$\overline{MS}$-NDR scheme can be found in \cite{Becirevic:2001xt}, and the decay constant of $B_s$ is from
\cite{Laiho:2009eu}. The relevant CKM matrix elements are obtained from the CKMfit collaboration
\cite{Charles:2004jd}. The SM Wilson coefficients for the quark  level
$b \rightarrow s \bar{q}^\prime q$ transition at next-to-leading order in the NDR scheme
are obtained from  \cite{Buchalla:1995vs}.

In the RS model, the matrix elements $M^{12,KK}_s$ and
$\Gamma^{12,KK}_s$ depend on the four couplings
$(U^{d(1)}_{L(R)})_{32}$, and $(U^{u(1)}_{L(R)})_{22}$ [see
Eq.~(\ref{RScoup})]. Writing these couplings explicitly yield%
\bea%
(U^{d(1)}_{L})_{32} &=&
V^{d\dagger}_{L(32)}V^{d}_{L(22)}[g^{(1)}_{L}(c_{Q_2})-g^{(1)}_{L}(c_{Q_1})]+
V^{d\dagger}_{L(33)}V^{d}_{L(32)}[g^{(1)}_{L}(c_{Q_3})-g^{(1)}_{L}(c_{Q_1})],\nl
(U^{d(1)}_{R})_{32} &=&
V^{d\dagger}_{R(32)}V^{d}_{R(22)}[g^{(1)}_{R}(c_{D_2})-g^{(1)}_{R}(c_{D_1})]+
V^{d\dagger}_{R(33)}V^{d}_{R(32)}[g^{(1)}_{R}(c_{D_3})-g^{(1)}_{R}(c_{D_1})],\nl
(U^{u(1)}_{L})_{22} &=&
V^{u\dagger}_{L(21)}V^{u}_{L(12)}[g^{(1)}_{L}(c_{Q_1})-g^{(1)}_{L}(c_{Q_3})]+
V^{u\dagger}_{L(22)}
V^{u}_{L(22)}[g^{(1)}_{L}(c_{Q_2})-g^{(1)}_{L}(c_{Q_3})],\nl
(U^{u(1)}_{R})_{22}&=& V^{u\dagger}_{R(21)}V^{u}_{R(12)}[g^{(1)}_{R}(c_{U_1})-g^{(1)}_{R}(c_{U_3})]+
V^{u\dagger}_{R(22)}
V^{u}_{R(22)}[g^{(1)}_{R}(c_{U_2})-g^{(1)}_{R}(c_{U_3})],
\label{explicitU}
\eea %
where the unitarity  of the  $V^{d(u)}_{L(R)}$ is used. The bulk
parameters $c_{f_\alpha}$ specify the position of the fermion's
localized wavefunction in the bulk. The specific choice of c's are
model dependent and they should generate the 4-d Yukawa hierarchy
of the quarks as well as the CKM mixing in the left-handed sector. In our numerical analysis, we do not consider
any specific values for c's.  We scan the allowed values for the
couplings $(U^{d(1)}_{L(R)})_{32}$, and $(U^{u(1)}_{L(R)})_{22}$
after imposing the constraints from the experimental measurements of
$\Delta M_s$ and $\Delta \Gamma_s$. Assuming the matrices $V^d_{LR}$ to be CKM-like and using $g^{(1)} \lsim 4$ one notes from
Eq. (\ref{explicitU}),  that  the coupling
$(U^{d(1)}_{L(R)})_{32}$ is constrained to $(U^{d(1)}_{L(R)})_{32} \lsim
0.1$, while $(U^{u(1)}_{L(R)})_{22}$ is of order one. Therefore, we
vary the four KK couplings as : $|(U^{d(1)}_{L(R)})_{32}| \lsim 0.1 $,  $|(U^{u(1)}_{L})_{22}| \lsim 1$, and
$|(U^{u(1)}_{R})_{22}| \lsim 3$, and their phases in the range [0, $2 \pi$].
 Note that this assumption of $U^d_L$ and $V^d_L$ being CKM-like is consistent with the scenario of anarchic 5D Yukawa couplings 
that has been considered in Ref.~\cite{ag}. Our choice for the right-handed mixing angles are not quite consistent with the scenario of anarchic 5D Yukawa couplings but our assumption is that with
proper tuning of the 5D Yukawa couplings we can  have the mixing angles in the range considered in this analysis.

In the fit, $\Delta M_s$ and $\Delta \Gamma_s$ are constrained by
 their experimental results within $1 \sigma$ errors. All SM
input parameters are uniformly varied within their errors and the SM
Wilson coefficients are evaluated at $\mu_b =4.2$ GeV. The bag
parameters are kept at their central values and $m_{KK}$ is varied in the range [1.2, 10.0] TeV.
Note that strong constraints on $m_{KK}$ are obtained from measurements in $K$ mixing \cite{K1} but there are still regions of parameter space (\cite{K2,Agashe:2008uz,K4}) which allow values for $m_{KK}$ considered in this work. Note that
the lower end of the KK scale which is scanned here (a couple of TeV)
would require some amount of tuning in order to satisfy the constraint
from $\epsilon_K$ (even with Higgs in the bulk)\cite{K2}.
Similarly, the KK scale of 1.2 to 2 TeV used in the scan is
generally disfavored by electroweak precision tests \cite{ew1, ew2}. Such
a low KK scale might be allowed if the 1st and 2nd generation fermions are
all chosen to have a (very) close-to-flat profile \cite{ew2}, but then one loses
the explanation of fermion mass hierarchies based solely on profiles.

\begin{table}
\caption{Numerical values of the theoretical quantities used in the numerical analysis are shown.}
 \centering
\begin{tabular}{|l|l|}
 \hline \multicolumn{2}{|c|}{Numerical values for the input parameters.
 }
 \\ \hline $m_{B_s}$ = 5.366 GeV & $f_{B_s}$= 238(9.5) MeV \\
 $m_b(m_b)$ = 4.19(+0.18)(-0.06)  GeV& $B_1$=0.87 \\
 $m_c(m_c)$ = 1.27(+0.07)(-0.09) GeV& $B_2$=0.84 \\
 $m_s$(2 GeV) = 0.01 GeV & $B_3$=0.91 \\
$m_s(m_b)$ =0.084 GeV & $B_4$=1.16 \\
 $\tau_{B_s}$ = 1.425 (0.041) ps  & $B_5$ = 1.75 \\
   $|V_{cb}|= 0.04128$  & $c_{Q_1}$ = 0.72 \\
  $|V_{cs}|$ = 0.97342 & $c_{Q_2}$ = 0.6 \\
  $|V_{tb}|$ = 0.999141  & $c_{Q_3}$ = 0.35 \\
  $|V_{ts}|$ = 0.04054 & $c_{U_1}$ = 0.63 \\
 $C_1(m_b)$=-0.1903 & $c_{U_2}$ = 0.30 \\
 $C_2(m_b)$= 1.081 & $c_{U_3}$ = 0.10 \\
 $C_3(m_b)$= 0.0137 & $c_{D_1}$ = 0.57 \\
 $C_4(m_b)$=-0.036 & $c_{D_2}$ = 0.57 \\
 $C_5(m_b)$= 0.009 & $c_{D_3}$ = 0.60 \\
  $C_6(m_b)$=-0.042 &  \\
   \hline
  \end{tabular}
  \label{tabinp}
\end{table}

\begin{figure}[htb!]
\centering
\includegraphics[width=7.5cm]{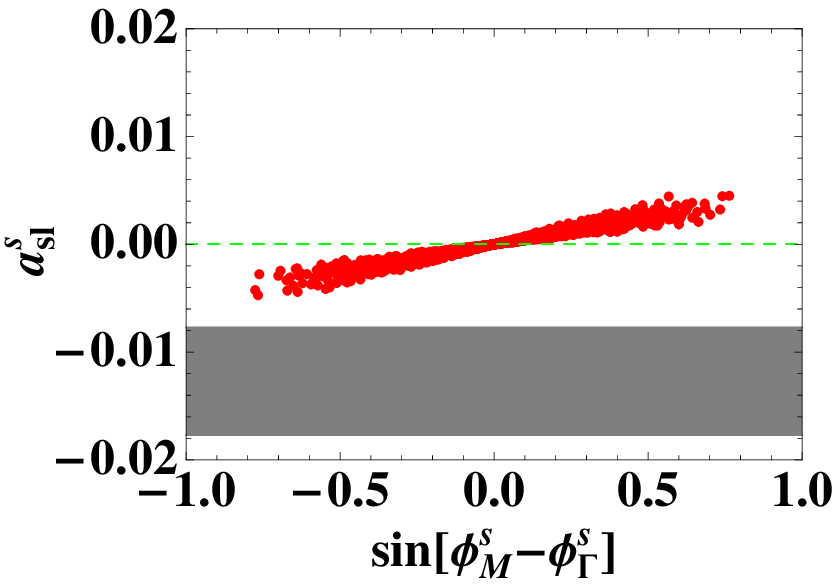}
\includegraphics[width=7.9cm]{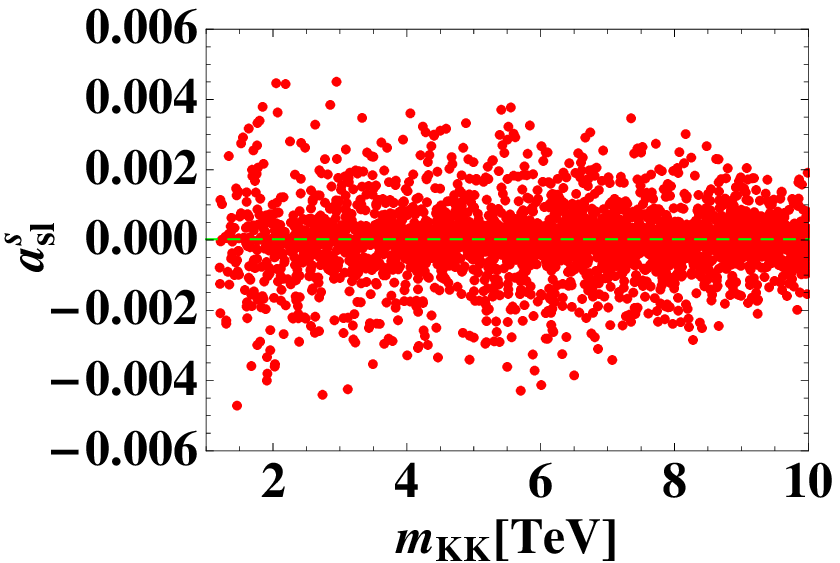}
\caption{ The  $a^s_{sl}$- $\sin{(\phi^s_M-\phi^s_\Gamma)}$ (left panel) and  $a^s_{sl}$- $ m_{KK}$ (right panel) correlation plots in the RS model. The gray band indicates the 1$\sigma$
 experimental allowed ranges for  $a^s_{sl}$ (avg). The green line indicates the SM prediction for $a^{s}_{sl}$.} \label{RSresults}
\end{figure}
We observe that  $\Delta M_s$(exp) constrains $|(U^{d(1)}_{L})_{32}|$ to $ \lsim 10^{-2}$, and allows  $r^{s,KK}_M =| M^{s,KK}_{12}/ M^{s,SM}_{12}| \lsim 1$. Note that similar conclusion about
$r^{s,KK}_M$ was also reached in Ref.~\cite{K2}. The fit results allow the ratio $r^{s,KK}_\Gamma =|
\Gamma^{s,KK}_{12}/ \Gamma^{s,SM}_{12}| \lsim 10 \%$ for values of $m_{KK}<$ 1.5 TeV and this ratio falls quite quickly as $m_{KK}$ is increased beyond 1.5 TeV. As indicated earlier, NP in $ b \to c \bar{c} s$ transitions will also contribute to the lifetimes of $b$ hadrons. We expect the corrections to the total widths of the $b$ hadrons to be also $\lsim 10 \%$ which cannot be detected experimentally given the hadronic uncertainties in calculating the total widths \cite{Lenz2}. The lifetime ratio of the $B_s$ meson to the $B_d$ meson, $\frac{\tau_s^{SM}}{\tau_d^{SM}}$, has a tiny theoretical uncertainty in the SM \cite{Lenz2}. New physics will contribute equally to the $B_s$ and $B_d$ total widths, in the leading order in the heavy quark expansion, thereby largely canceling in the lifetime ratio. Our naive estimate is, NP in $ b \to c \bar{c} s$ transition can correct this lifetime ratio as 
$\frac{\tau_s}{\tau_d} \approx \frac{\tau_s^{SM}}{\tau_d^{SM}}(1+x)$, with
$ x=\frac{X}{\Gamma_d^{SM}} \left( 1 - \frac{\Gamma_d^{SM}}{\Gamma_s^{SM}} \right) $ where $\Gamma_{s,d}^{SM}$ are the SM $B_{s,d}$ widths. Using $\frac{X}{\Gamma_d^{SM}} \lsim $ 10 \%,
 $\frac{\tau_s^{SM}}{\tau_d^{SM}} = 1 \pm 0.01$ \cite{Lenz2}, we get
 $x \lsim 0.1 \%$ which is consistent with experimental measurements \cite{pdg}. In  Fig. \ref{RSresults} is shown the  $a^{s}_{sl}$ - $\sin{(\phi^s_M-\phi^s_\Gamma)}$  and  $a^{s}_{sl}$-$m_{KK}$  correlation plots in the RS model. The gray band indicates the 1$\sigma$  experimental allowed ranges for  $a^s_{sl}$ (avg) while the green line indicates the SM prediction for $a^{s}_{sl}$.  As one can see from these figures, the fit results allow  $a^s_{sl} \sim  -0.00498$, which is $1.54 \sigma $ away from its experimental average value in 
 Eq.(\ref{asls.avg}) . The value of the corresponding $\sin {(\phi_M - \phi^s_\Gamma )}$ is $-0.76$.
 Hence this model cannot fully account for the experimental results and this is due to the fact that this model cannot generate enough correction to the width difference $\Delta \Gamma_{12}^s$.
Note that, as $m_{KK}$ is increased the suppression to $a^s_{sl}$ due to $m_{KK}$ is partially compensated by larger mixing angles which are within the considered ranges in this work and are  consistent with experimental measurements of $\Delta M_s$ and $\Delta \Gamma_s$. However, after a certain point the
 suppression due to $m_{KK}$ cannot be compensated and $a^s_{sl}$ decreases with increasing $m_{KK}$ mass.

\section{Conclusions}
In the past few years several measurements in rare B decays
involving $ b \to s$ transitions have been somewhat difficult to
understand in the SM. This has put the focus on measurements in
the $\bs-\bsbar$ system. The measurements of the phase in
$\bs-\bsbar$ mixing and more recently the like-sign dimuon
asymmetry have generated an enormous interest in the community as
these measurements indicate possible deviations from the SM
predictions. In this work, we presented calculations for general
new physics corrections to the parameters in $\bs-\bsbar$ mixing.
Taking the Randall-Sundrum model as an example of new physics we
calculated the  wrong-charge asymmetry, $a_{sl}^s$, as well as other
parameters in  $\bs-\bsbar $ mixing. Our calculations indicate
that while the RS model can cause deviations from the SM
predictions for the wrong-charge asymmetry, it cannot explain
the present experimental average value within the 1$\sigma$ range.
This is due its inability to generate
sufficient new contribution to the width difference $\Delta \Gamma^s_{12}$, even though the model can generate large contribution to the mass difference  $\Delta M^s_{12}$.

\section{Acknowledgements}
We would like to thank Q. Shafi, Y. Mimura B. Dutta and Alexander Lenz for useful discussions. The work of S. K. was partially supported by
the Science and Technology Development Fund (STDF) Project ID
1855, the ICTP Project ID 30, and the Academy of Scientific
Research and Technology. The work of A.D and M.D  was supported by the U.S.-Egypt Joint  Board on Scientific and Technological Cooperation through the U.S. Department of Agriculture.


\end{document}